\documentclass[twocolumn,superscriptaddress,showpacs,prl,floatfix]{revtex4}
\usepackage{graphics,amsmath,graphicx}
\usepackage{grffile}
\usepackage{color}

\newcommand{\pbste}{PbS$_x$Te$_{1-x}$ }
\newcommand{\pbsnte}{Pb$_{1-x}$Sn$_x$Te }

\begin{document}

\title{Robustness against Disorder of Relativistic Spectral Properties in
Chalcogenide Alloys}

\author{Domenico Di Sante} \affiliation{Consiglio Nazionale delle Ricerche
(CNR-SPIN), Via Vetoio, L'Aquila, Italy} \affiliation{Department of Physical and
Chemical Sciences, University of L'Aquila, Via Vetoio 10, I-67010 L'Aquila,
Italy}\email{domenico.disante@aquila.infn.it}

\author{Paolo Barone} \affiliation{Consiglio Nazionale delle Ricerche
(CNR-SPIN), Via Vetoio, L'Aquila, Italy}

\author{Evgeny Plekhanov} \affiliation{Consiglio Nazionale delle Ricerche
(CNR-SPIN), Via Vetoio, L'Aquila, Italy}

\author{Sergio Ciuchi} \affiliation{Department of Physical and Chemical
Sciences, University of L'Aquila, Via Vetoio 10, I-67010 L'Aquila, Italy}
\affiliation{Consiglio Nazionale delle Ricerche (CNR-ISC), Via dei Taurini,
Rome, Italy}

\author{Silvia Picozzi} \affiliation{Consiglio Nazionale delle Ricerche
(CNR-SPIN), Via Vetoio, L'Aquila, Italy}

\date{\today}

\begin{abstract}

In order to carefully address the interplay between substitutional disorder and
spin-orbit-coupling in IV-VI  alloys, we propose a novel theoretical
approach that integrates the reliability of plane-wave based density-functional
theory beyond the local-density approximation with the Coherent Potential
Approximation. By applying the proposed method to ternary chalcogenide alloys, we
predict a substantial robustness of spectral features close to the Fermi energy
against substitutional disorder. Supplementing our first-principles calculations
with the analysis of the  $k \cdot p$ model for rock-salt chalcogenides, we show
that the disorder self-energy is vanishingly small close to the band gap, thus
allowing for bulk Rashba-like spin splitting to be observed in ferroelectric alloys,
such as PbS$_x$Te$_{1-x}$, and protecting the band-character inversion related
to the topological transition in the recently discovered Topological Crystalline
Insulator Pb$_{1-x}$Sn$_x$Te.

\end{abstract}


\maketitle

{\bf Introduction.} 
It is nowadays well accepted that  spin-orbit coupling (SOC) lies at the origin
of a rich variety of appealing phenomenologies, ranging from topological
insulators (TI) \cite{TI1,TI2}, showing fully spin-polarized metallic surface
states despite their bulk insulating character, to large bulk Rashba-like
spin-splitting effects in noncentrosymmetric and polar materials
\cite{BiTeI1,BiTeCl,GeTe}, characterized by strong spin-momentum locking and
spin-polarized bands. In this respect, the long-known class of main-group IV-VI
semiconducting chalcogenides is playing a key role. On one hand, SnTe has been
proposed\cite{SnTeTCI}, and later confirmed\cite{SnTe_exp}, to belong to a new
class of TI with unique features arising from the combination of time-reversal
and crystalline symmetries, hence named Topological Crystalline Insulators
(TCIs)\cite{Fu11,Slager13}. On the other hand, GeTe --- one of the few known
ferroelectric semiconductors --- has been put forward as the first example of
so-called Ferroelectric Rashba Semiconductors (FERSCs) with large Rashba
splitting\cite{SilviaFERSC}, where the possibility to permanently control its
spin texture via a switchable electric polarization could introduce new
functionalities in spintronic devices\cite{GeTe}. In principle, bulk Rashba-like
effects and topological features could even be realized within the same
material. As a matter of fact, SnTe itself  undergoes a low-temperature
ferroelectric (FE) transition under certain conditions\cite{kobayashi.prl1976},
whose potential implications on its TCI phase have been recently theoretically
investigated\cite{SnTe_FE}. The potential relevance of noncentrosymmetric TI
with Rashba-like spin-splitting has been also proposed for the acentric
semiconductors BiTeI under pressure\cite{BiTeI3} and BiTeCl\cite{BiTeCl}. 

Beside GeTe and SnTe, FE transitions have been reported in several ternary or
quaternary alloys of rock-salt IV-VI chalcogenides, such as
Pb$_{1-x}$Ge$_{x}$Te, Pb$_{1-x}$Sn$_{x}$Te and
PbS$_x$Te$_{1-x}$\cite{Lines77,PbSTe1,PbSTe2}. Similarly, TCI features have been
observed in photoemission (ARPES) spectra of n-type Pb$_{1-x}$Sn$_x$Te and
p-type Pb$_{1-x}$Sn$_x$Se  alloys as a function of doping \cite{ARPES_PbSnTe,
Tanaka.prb2013, ARPES_PbSnSe, Wojek.prb2013, Polley.prb2014, wojek.arxiv2014}.
When addressing the relativistic properties of such alloys,  
due to the intrinsically disordered character of the solid-state
solution, a major issue
appears. While the role of disorder in $\mathcal{Z}_2$ TI has been extensively
investigated\cite{Z2DIS1,Z2DIS2,Z2DIS3}, showing that time-reversal symmetry
guarantees the presence of metallic edge states, topologically protected from
non-magnetic impurities,  it is not obvious how disorder entangles with
spin-orbit  in TCI alloys, since crystalline symmetries, crucial for topological
properties, are  locally broken in a random way. On the basis of general
arguments, delocalized states were shown to survive at the surface of TCIs as
long as the relevant symmetries, which are broken by disorder, are restored by
disorder averaging, analogously to what happens in weak $\mathcal{Z}_2$
TI\cite{Fu_disorder.prl2012}. However, signatures of disorder should manifest in
the ARPES spectra\cite{wojek.arxiv2014}. Similarly, broadening/smearing effects
induced by disorder could in principle screen, reduce or even prevent any bulk
Rashba-type spin-splitting in FE alloys.

In this Letter, we investigate the role of disorder in the relativistic
electronic properties of IV-VI chalcogenide ternary alloys by interfacing a
Coherent-Potential-Approximation (CPA) approach\cite{CPA} to highly accurate
density-functional theory (DFT) relativistic calculations. Aiming at a
quantitative analysis of the effect of disorder in the spectral features of such
alloys, we calculate the disorder self-energy in the CPA framework, using, as a
starting point, accurate DFT calculations on ordered binary compounds based on
the HSE hybrid functional, as implemented in  VASP \cite{HSE,VASP,DFT_INFO}.
Demanding hybrid-functional calculations have been shown to provide, in good
agreement with GW calculations\cite{Hummer, Cardona}, the correct band ordering
at the valence band maximum (VBM) and conductance band minimum (CBM) in all lead
chalcogenides, at variance with severe failures within local-density
approximation. This issue is of dramatic importance in rock-salt chalcogenides,
where a band-character inversion at the high-symmetry L point drives the
topological transition, with a change in the mirror Chern number by
two\cite{SnTeTCI,Slager13}. Ab-initio relativistic HSE band structures are then
projected onto a Wannier basis set \cite{WANNIER90}, and Wannier-functions
tight-binding parameters are used for the CPA self-consistency loops (see
details in \cite{SuppMat}). We stress that, to the best of our knowledge, this
is the first time that such hybrid DFT-CPA approach is used in the study of
disordered systems, hopefully providing an alternative approach to
well-established techniques, such as the fully relativistic
KKR-CPA\cite{MinarKKR,Felser,Stocks}, often based on local-density
approximation. Our approach also allows to tailor disorder effects in the
surface states of alloys without resorting to computationally demanding DFT
supercell calculations, at the same time improving on the conventional
methodology, where tight-binding Hamiltonians are built up in slab geometry. In
the latter approach, substitutional disorder in TCI chalcogenide alloys has been
so far taken into account within the Virtual Crystal Approximation
(VCA)\cite{ARPES_PbSnSe}, where no self-energy correction accounting for the
effective disorder interaction is included. On the other hand, the spectral
features of the surface states are encoded in the Surface Single-site Green's
Function, which can be easily calculated from the DFT-CPA single-site
disorder-averaged Green's Function  using an iterative surface renormalization
method in the limit of a semi-infinite slab\cite{HenkSurface,SuppMat}.

On the grounds of our newly developed metholodology, we focus on two ternary
alloys, \pbste and \pbsnte, showing that the former belongs to the class of
FERSC, where a large Rashba-like spin-splitting develops in  disordered FE
alloys, while we provide a microscopic analysis of the topological nature of the
TCI phase that develops in the latter.

{\bf Application to PbSTe FERSC alloy.} We first focus on
the interplay between disorder and Rashba-type spin splittings in 
FE PbS$_x$Te$_{1-x}$  \cite{PbSTe1,PbSTe2}. 
The pristine compounds PbTe and PbS don't display any FE instability, both crystallizing in the centrosymmetric face-centered cubic structure.  
On the other hand, their alloys, up to a sulfur concentration of $x_c\sim 0.45$, show FE
phase transitions at doping-dependent Curie temperatures, displaying a maximum $T_c\sim 80~K$ around $x=0.18$\cite{PbSTe2}.
\begin{figure}[!hb]
\centering
\includegraphics[width=0.47\textwidth,angle=0,clip=true]{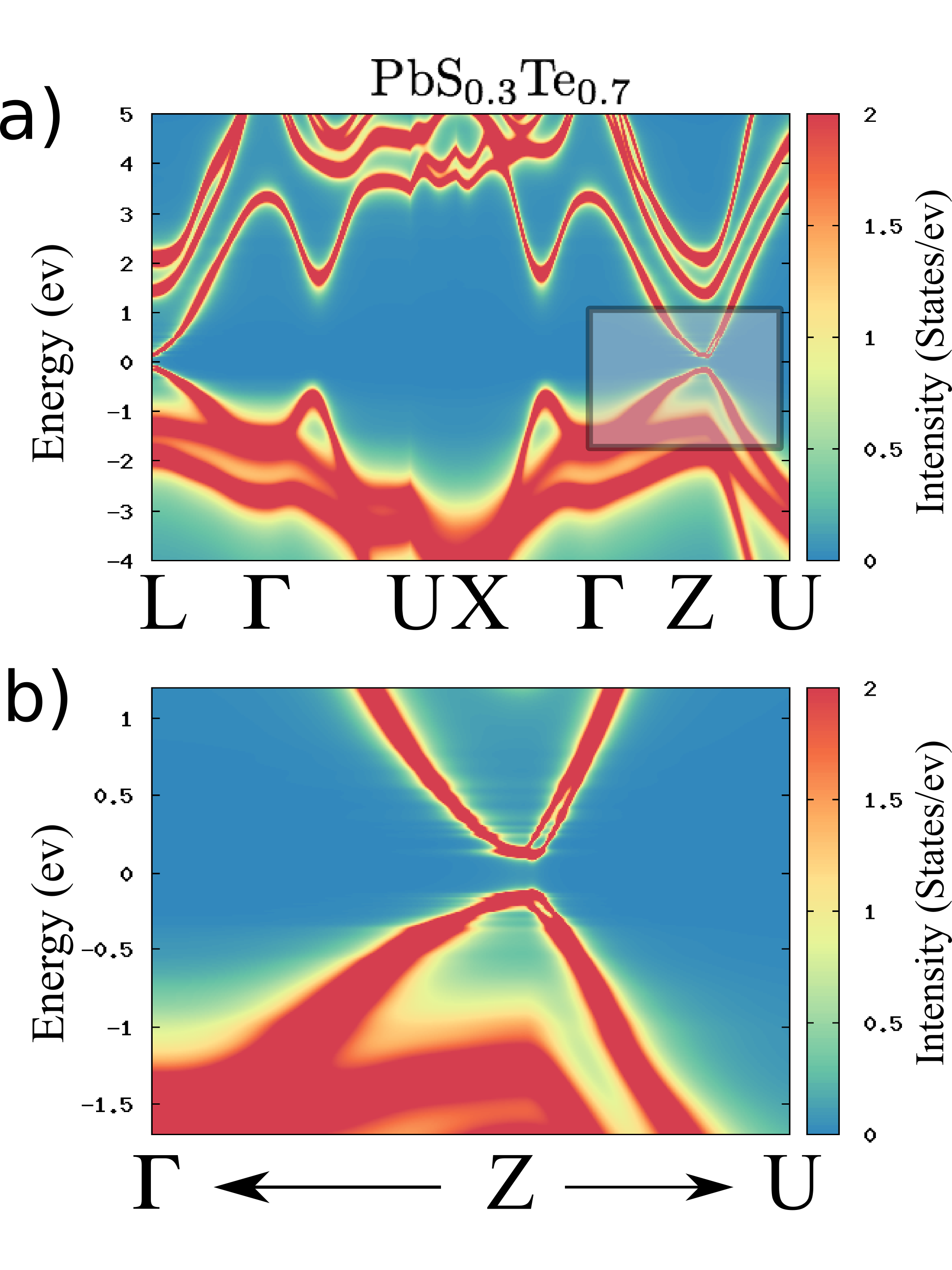}
\caption{(Color online) (a) DFT(HSE)+CPA Spectral Function for PbS$_{0.3}$Te$_{0.7}$
FE alloy along the full irreducible rhombohedral BZ, where a clear spin-splitting signature appears in the black rectangle. (b) Zoom of the spectral function around the Z point, corresponding to the black rectangle in panel (a), highlighting a Rashba-like
induced spin-splitting of about 50 $meV$ along the ZU line perpendicular to the polar axis.
} 
\label{fig:PbSTe}
\end{figure}
\begin{figure*}[!ht]
\centering
\includegraphics[width=0.70\textwidth,angle=0,clip=true]{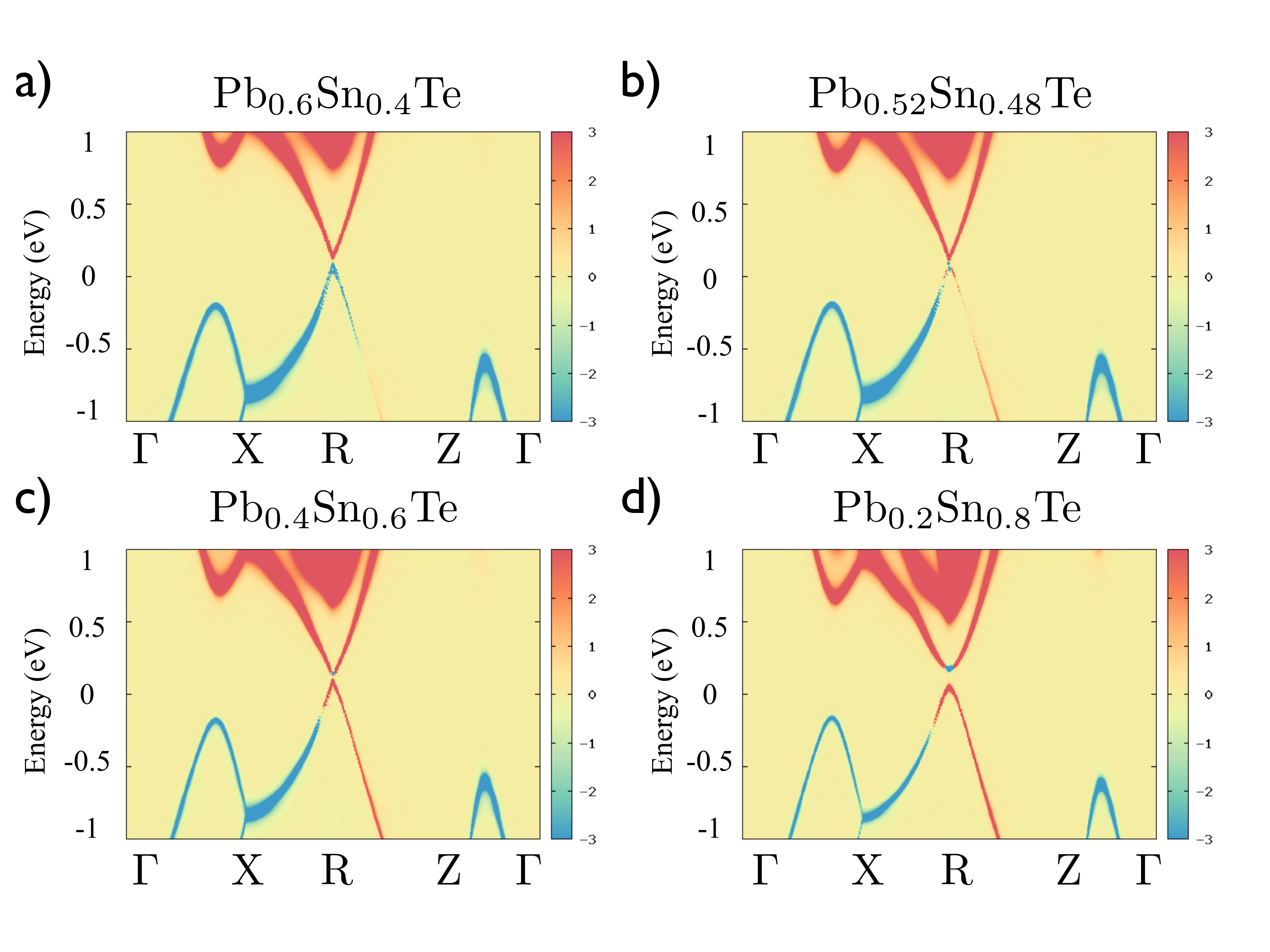}
\caption{(Color online) DFT(HSE)+CPA evolution of Pb$_{1-x}$Sn$_x$Te alloy
spectral features as a function of Sn concentration for x=0.4, 0.48, 0.6
and 0.8 in a),b),c) and d) panels respectively, along the irreducible tetragonal
BZ. Red (positive) values refer to the spectral function $ A({\bf
k},\omega)=- {\text Im} {\text Tr} {\cal G} ({\bf k},\omega)/\pi $ projected on the
 Pb$_{1-x}$Sn$_x$ cation, while blue (negative) values are projection of the
spectral function on the  Te anion. The R point is the projection of the
rhombohedral L point on the tetragonal BZ.} 
\label{fig:TCIbulk}
\end{figure*}
By means of EXAFS, the origin of such transition has been ascribed to the
off-centering of S ions in a substantially centrosymmetric PbTe
host\cite{PbSTe1}, due to different size and polarizability of S and Te ions.
Our DFT supercell calculations confirm this scenario, finding that below a
critical concentration $x_c\sim 0.4$ sulfur displacements vary in the same range
that has been experimentally observed, whereas only Te ions close to S are
slighlty displaced (see details in \cite{SuppMat}).

At the FE transition, the alloy structure changes from cubic to distorted
rhombohedral (space group $R3m$), being thus isostructural with the prototypical
Ferroelectric Rashba Semiconductor GeTe\cite{GeTe}. 
Rashba-like spin splittings are therefore expected
to develop in the presence of SOC, especially at the VBM and CBM around the
high-symmetry point Z, at the center of the hexagonal faces in the Brillouin
zone (BZ) which are perpendicular to the [111]  polar axis\cite{GeTe}, provided
disorder-induced broadening effects do not overcome such tiny features of the
spectral function. Indeed, strong signatures of the random S/Te substitution
appear, especially in the valence bands with predominant anionic character, as
shown in Fig. \ref{fig:PbSTe}a). Nonetheless, {\em the smearing of the spectral
function is substantially reduced around the Fermi level}. Remarkably,  we find
that the VBM and CBM are more robust against disorder renormalization as the
band gap decreases. In fact, the disorder self-energy appears to depend not only
on the nominal S/Te concentration, but also on the size of the gap, which in
turn is loosely proportional to the amount of S off-centering. Since the latter
decreases as the S concentration increases above $x\sim0.18$, spin-splitting
signatures become clearly visible due to the smaller disorder self-energy, as
shown in Fig. \ref{fig:PbSTe}b) for $x=0.3$. On the other hand, for small S
concentrations such splitting effects are substantially depressed, virtually
disappearing, by disorder renormalization effects, despite the nominally weaker
substitutional disorder. We stress the fact that our DFT+CPA approach is
essential in capturing such disorder-induced renormalization effects. As a
matter of fact, a VCA approach --- not taking into account the disorder
self-energy --- would predict large spin splittings for all concentrations
corresponding to the FE phase (see details in Suppl. Mat. \cite{SuppMat}). Being
isostructural with FE GeTe and SnTe, our results suggest that novel FERSC
materials, allowing for the control and manipulation of Rashba-like spin
textures by tuning the FE polarization\cite{GeTe,SilviaFERSC}, could be
identified in the broad class of chalcogenide alloys. Depending on doping
concentration, tiny features of ARPES spectra could become visible in \pbste or
related PbSe$_x$Te$_{1-x}$\cite{PbSTe2}, as the self-energy induced by
substitutional disorder may be significantly reduced close to the Fermi energy.

{\bf Application to PbSnTe TCI alloy.} We turn now to the study of the well
known topological transition as a function of tin  concentration in the
Pb$_{1-x}$Sn$_x$Te alloy. This compound, along with the analogue
Pb$_{1-x}$Sn$_x$Se, has been intensively studied, both experimentally and
theoretically \cite{ARPES_PbSnTe, Tanaka.prb2013, ARPES_PbSnSe, Wojek.prb2013,
Polley.prb2014,wojek.arxiv2014}. For the first time, we propose in this Letter
an ab-initio analysis of the disorder effects on its spectral features by
calculating the disorder self-energy in our combined DFT-CPA framework. In Fig.
\ref{fig:TCIbulk}, we show the anion/cation resolved spectral functions for
different dopings. The CPA approach clearly describes the doping-induced band
closure at the L points (R points in the tetragonal setting), implying a change
of the alloy topological character, with a critical tin concentration $x_c\sim
0.48$ \cite{Note_Xc}. As a consequence of the cation Pb/Sn substitution,
disorder effects are more visible at positive energies with respect to the Fermi
level, i.e. where states have a predominant cationic orbital character. The
presence of disorder is reflected in the significant broadening of the spectral
features (spectra in wider energy windows are reported in Supplementary
Material\cite{SuppMat}), as a consequence of a larger self-energy. However,
approaching the Dirac point --- where bands show an almost linear behavior ---
the spectral broadening vanishes, and a clear signature of character inversion
is visible. Remarkably, such inversion appears to be protected against disorder,
since the imaginary part of the disorder self-energy is found to vanish. As a
result, {\em bands are not renormalized (broadened) by the interaction with
disorder around the bulk Dirac point}.

\begin{figure}[!ht]
\centering
\includegraphics[width=0.40\textwidth,angle=0,clip=true]{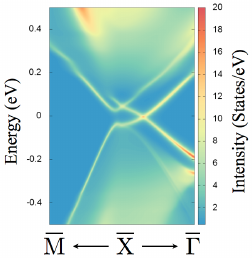}
\caption{(Color online) DFT(HSE)+CPA Surface Spectral Function for
Pb$_{0.2}$Sn$_{0.8}$Te TCI alloy (see Fig. \ref{fig:TCIbulk}d for the bulk
spectral function). The k-axis show 1/10 of the ${\bar {\text M}}-{\bar {\text
X}}$ and ${\bar {\text X}}-{\bar \Gamma}$ lines around ${\bar {\text X}}$. } 
\label{fig:TCIsurface}
\end{figure}

The trivial/topological nature of the bulk bandstructure is reflected in the
lack/presence of metallic surface states through the bulk-boundary
correspondance principle\cite{TI2}. In particular, an inverted orbital
cation/anion character in the rock-salt chalcogenides ensures the appearance of
Dirac-like surface states\cite{SnTeTCI}. In fact, we find that gapless surface
states appear as the tin concentration overcomes the critical value $x_c\sim
0.48$ (for a trend of the surface spectral function as a function of doping see
the Supplementary Material \cite{SuppMat}). The case of Pb$_{0.2}$Sn$_{0.8}$Te
TCI alloy is shown in Fig. \ref{fig:TCIsurface}, where a surface Dirac cone
located close to the ${\bar {\text X} }$ (as in pristine SnTe\cite{SnTeTCI}) is
clearly visible. 
As it happens in the bulk band-structure, surface states appear to be well
protected against disorder-induced broadenings around the Dirac point. We
emphasize that surface spectral functions calculated within our approach find a
direct link with experimental ARPES spectra, apart from matrix element effects
(see e.g. Refs\cite{ARPES_PbSnTe, ARPES_PbSnSe,wojek.arxiv2014}).

{\bf CPA on the $k\cdot p$ model for TCI.} The origin of the Dirac-point
protection against substitutional disorder can be further analyzed by
considering the CPA description of a trivial/topological alloy in the framework
of $k\cdot p$ model. The band structure of rock-salt chalcogenides around the L
points is given by\cite{SnTeTCI,MitchellWallis}:
\begin{eqnarray} 
\label{form1} 
\hat{H}^\pm = \pm m\sigma_z + \nu(k_1s_2 - k_2s_1)\sigma_x + \nu'k_3\sigma_y \quad, 
\end{eqnarray}
where ${\mathbf k}=(k_1,k_2,k_3)$ forms an orthogonal system with $k_3$ parallel
to the $\Gamma {\text L}$ direction and $k_1$ along the direction perpendicular
to a mirror plane. Hamiltonian (\ref{form1}) is represented in the basis of
cation and anion p-orbitals p$_c$/p$_a$ ($\sigma_z = \pm 1$) and total angular
momentum along $\Gamma {\text L}$ $j=\pm 1/2$ ($s_3 = \pm 1$). $\nu$ and $\nu'$
are materials-dependent parameters. The plus/minus sign in front of the mass
term $m$ refers to the trivial/topological insulator with direct/inverted band
character at the gap\cite{SnTeTCI}.  The CPA description of the
trivial/topological material alloy is based on the single-site
disordered-averaged propagator: 
\begin{eqnarray} 
\label{form2}
{\cal \hat{G}}(\omega) &=& x[\hat{G}_0(\omega)^{-1}-\hat{H}^+]^{-1}  \nonumber \\
                      &+& (1-x)[\hat{G}_0(\omega)^{-1}-\hat{H}^-]^{-1} \quad , 
\end{eqnarray} 
with $\hat{H}^{\pm}$ being the local (k-independent) part of Hamiltonian
(\ref{form1}) and $\hat{G}_0$ being the
local propagator which embodies the average action of the environment. The disorder self-energy ($\hat{\Sigma}(\omega)$) defined as
\begin{eqnarray} 
\label{form3}
{\cal \hat{G}}(\omega) = [\hat{G}_0(\omega)^{-1}-\hat{\Sigma}(\omega)]^{-1} 
\end{eqnarray} 
contains all information relative to the role played by
disorder. All the detailed CPA calculations are extensively given in the
Supplementary Material \cite{SuppMat}; here we will focus only on the most
relevant results. At the critical concentration $x_c$ where the bulk gap closes
with bands linearly dispersing, the disorder self-energy can be written as
$\Sigma(\omega)=-a\omega -ib\omega^2$ with $a\sim m^2/\Gamma^2$ and $b\sim
m^2/\Gamma^3$, where $\Gamma$ is a high energy cut-off naturally introduced when
dealing with linearized models\cite{Note_confdis}. The density of states 
$N(\omega)=-{\text Im} {\cal G}(\omega)/\pi$ assumes the simple parabolic expression
$N(\omega)=3\omega^2/2\Gamma^3$ coming from the tridimensionality of the Dirac
cone. Remarkably, the imaginary part of the disorder self-energy, which is
related to the finite life-time and bandwidth of electronic states, goes
monotonically to zero when approaching the Dirac point
$(\omega=0)$, vanishing at the Fermi level. This is of fundamental importance
for the anion/cation band character inversion related to the topological
transition\cite{SnTeTCI}, since the energy region of interest for the orbital
character inversion around the L points is protected against disorder-induced
broadening effects. Interestingly, such a protection is absent in graphene,
where a non vanishing self-energy at the Fermi level leads to the so-called
universal conductivity (see details in \cite{SuppMat}). It is also worth to note
here that our results for the 3D Dirac cone at the trivial/topological
transition show many similarities with those obtained for Weyl electron
systems and three dimensional $\mathcal{Z}_2$ topological insulators in the weak
conformational disorder limit \cite{Ominato,sbierski.arxiv2014,Kobayashi}. On
the other hand, it is important to highlight that finite-range potential
disorder, as well as spatial correlations, beyond CPA, taken into account by
Dynamical Cluster Approximation, can lead to an exponentially small density of
states at the Dirac point, as a direct consequence of a small, but finite,
disorder self-energy at the Fermi level \cite{Huse,DCA}.

{\bf Conclusions:} In this work we've addressed the crucial role of substitutional
disorder in the class of IV-VI chalcogenides,  featuring SnTe and GeTe as   
binary prototypes of TCIs and FERSC, respectively. The main outcome
of our theoretical analysis, based on the CPA
model and ab-initio calculations, is the prediction that 
spectral features in proximity to the Fermi level are robust with respect
to substitutional disorder. In closer detail,  we predict a strong interplay between disorder
and Rashba spin-splittings in ferroelectric PbS$_x$Te$_{1-x}$ ternary
alloys. Moreover, in Pb$_{1-x}$Sn$_x$Te we find  that,   as the disorder-induced self-energy  vanishes
at the bulk Dirac point, the orbital character
inversion - ``smoking gun" of the topological transition - is protected
against the effects of disorder. As a result, the spectral features of the Dirac-like conducting surface states 
 are robust and therefore clearly detectable in experiments.

We kindly acknowledge A. Narayan and K. Palotas for useful discussions.
We acknowledge CINECA for awarding us access to
Fermi supercomputer and the CARIPLO Foundation through the MAGISTER project Rif. 2013-0726.

\bibliographystyle{apsrev}
\bibliography{biblio}

\begin{thebibliography}{46}
\bibitem[{DFT()}]{DFT_INFO}
\bibinfo{note}{{\text Kohn-Sham equations were solved using the projector
  augmented-wave method. The energy cutoff for the plane-wave expansion was 600
  eV; an 8x8x8 Monkhorst-Pack k-point grid was used.}}
\bibitem[{Not({\natexlab{a}})}]{Note_Xc}
\bibinfo{note}{{\text The agreement with experiment can be further improved
  performing the hybrid functional calculations not at the PbTe and SnTe
  equilibrium lattice constant but at the lattice constant of the alloy.}}
\bibitem[{Not({\natexlab{b}})}]{Note_confdis}
\bibinfo{note}{{\text Inclusion of a small gaussian conformational disorder
  with variance $\sigma<<m$ causes only a renormalization of coefficients $a$
  and $b$, where $m$ is replaced with $(m^2+\sigma^2)$\cite{SuppMat}.}}
\end{thebibliography}

\end{document}